# Pressure-induced huge increase of Curie temperature of the van der Waals ferromagnet $VI_3$

J. Valenta[1,2], M. Kratochvílová[1], M. Míšek[3], K. Carva[1], J. Kaštil[3], P. Doležal[1], P. Opletal[1], P. Čermák[1], P. Proschek[1], K. Uhlířová[1], J. Prchal[1], M. J. Coak[4,5], S. Son[6,7,8], J-G. Park[6,7,8], and V. Sechovský[1]

*[1]Charles University, Faculty of Mathematics and Physics, Department of Condensed Matter Physics, Ke Karlovu 5, 121 16 Prague 2, Czech Republic*
*[2]National Institute for Materials Science, Thermal Energy Materials Group, International Center for Materials Nanoarchitectonics (MANA), 1-2-1, Sengen, Tsukuba, Ibaraki 305-0047, Japan*
*[3]Institute of Physics, Czech Academy of Sciences, Na Slovance 2, 182 21 Prague 8, Czech Republic*
*[4]Department of Physics, University of Warwick, Gibbet Hill Road, Coventry CV4 7AL, UK*
*[5]Cavendish Laboratory, Cambridge University, J.J. Thomson Ave, Cambridge CB3 0HE, UK*
*[6]Center for Quantum Materials, Seoul National University, Seoul 08826, Republic of Korea*
*[7]Center for Correlated Electron Systems, Institute for Basic Science, Seoul 08826, Korea*
*[8]Department of Physics and Astronomy, Seoul National University, Seoul 08826, Korea*

**ABSTRACT**

Evolution of magnetism in single crystals of the van der Waals compound $VI_3$ in external pressure up to 7.3 GPa studied by measuring magnetization and ac magnetic susceptibility is reported. Four magnetic phase transitions, at $T_1$ = 54.5 K, $T_2$ = 53 K, $T_C$ = 49.5 K, and $T_{FM}$ = 26 K, respectively have been observed at ambient pressure. The first two have been attributed to the onset of ferromagnetism in specific crystal-surface layers. The bulk ferromagnetism is characterized by the magnetic ordering transition at Curie temperature $T_C$ and the transition between two different ferromagnetic phases $T_{FM}$, accompanied by a structure transition from monoclinic to triclinic symmetry upon cooling. The pressure effects on magnetic parameters were studied with three independent techniques. $T_C$ was found to be almost unaffected by pressures up to 0.6 GPa whereas $T_{FM}$ increases rapidly with increasing pressure and reaches $T_C$ at a triple point at ≈ 0.85 GPa. At higher pressures, only one magnetic phase transition is observed moving to higher temperatures with increasing pressure to reach 99 K at 7.3 GPa. In contrast, the low-temperature bulk magnetization is dramatically reduced by applying pressure (by more than 50% at 2.5 GPa) suggesting a possible pressure-induced reduction of vanadium magnetic moment. We discussed these results in light of recent theoretical studies to analyze exchange interactions and provide how to increase the Curie temperature of $VI_3$.

## INTRODUCTION

Two-dimensional van der Waals (vdW) magnetic materials have become the subjects of intensive research activities in recent years, mainly because of their promising application potential for the design of spintronic devices[1-4]. One of these intriguing vdW materials, $VI_3$, received significant attention from experimentalists[5-14] and theorists[15-27] only recently, despite belonging to the well-studied family of transition metal trihalides. It was stimulated by the break-through discovery of 2D ferromagnetism in $CrI_3$ monolayers[28], whose anisotropy allows us to overcome the effect of the Mermin-Wagner theorem. The studies on $VI_3$ rapidly appearing in a short time brought numerous surprising experimental results, especially on crystal structures, magnetic phase transitions, and their evolution in external pressure. Most experimental works agreed on the structural transition from the trigonal symmetry to a monoclinic one upon cooling through the transition temperature $T_s \approx 79$ K[5, 6, 8-10]. Such a transition from a higher to a lower crystal symmetry with decreasing temperature contrasts with the behavior of $CrI_3$, which has a monoclinic structure at room temperature. It is then transformed into a trigonal structure when cooled below 220 K[29]. In $VI_3$, the consensus is on the magnetic phase transition from paramagnetic (PM) to a ferromagnetic (FM) state at Curie temperature $T_C \approx 50$ K. A strong anisotropy with a high anisotropy field (> 9T at 2K) leads to a high coercive field $\mu_0 H_c \approx 1$ T at 2 K in a magnetic field applied parallel to the $c$*-axis[5-7, 10, 11]. It contrasts with a much lower anisotropy field (≈ 3 T at 2 K) and soft ferromagnetism observed in $CrI_3$[29].

Subsequent studies revealed that magnetism in $VI_3$ is more complicated than initially believed. Gati, et al.[8] reported that $VI_3$ undergoes another magnetic phase transition at 36 K (referred to as ≈ $T_{FM2}$) between two ferromagnetically ordered phases. Two magnetically ordered V sites were detected at the lowest temperatures ($T < 36$ K) while there is only one magnetically ordered V site at temperatures between 36 K and $T_C$ by $^{51}V$ and $^{127}I$ NMR spectroscopy[7]. The low-temperature X-ray diffraction indicated that this transition is associated with the reported transition between the monoclinic and triclinic structure at 32 K upon cooling[9].

Measurements of magnetic parameters of materials exposed to external pressure provide valuable information about the character of exchange interactions and magnetism's dimensionality. The application of hydrostatic pressure leads to reducing the interatomic distances in proportion to the corresponding direction's compressibility. Consequently, exchange integrals, the hierarchy of exchange interactions, and magnetic moments are modified, and the critical parameters like $T_C$ are continuously tuned by varying applied pressure. Natural materials' compressibility is often anisotropic, particularly in layered vdW crystal structures with different intralayer and interlayer bonding types. Two-dimensional vdW materials can serve as excellent examples of anisotropic compressibility under hydrostatic pressure, as it is several times larger within the parallel van der Waals bonds than in perpendicular directions[30-34]. When sufficient pressure is applied in a vdW material, usually an insulator, it may undergo a Mott insulator-metal transition. Besides a dramatic change of electrical conductivity, this transition is usually accompanied by a significant change in magnetic behavior[35, 36].

In $VI_3$, the first measurements of magnetic parameters in hydrostatic pressure[5] showed that $T_C$ remains almost intact in low pressures up to 0.6 GPa. The increase of $T_C$ at 1 GPa has been ascribed to tuning the $VI_3$ dimensionality away from 2D. The subsequent detailed measurements of temperature dependences of specific-heat and magnetization under various pressures[8] shed new light on the unusual pressure dependence of $T_C$. This study confirmed that $T_C$ (assigned in Ref. 8 as $T_{FM1}$) is almost unaffected by hydrostatic pressure up to 0.6 GPa and revealed that $T_{FM2}$ increases dramatically with increasing pressure, so that related transition merges with the FM transition at $T_C$ in the pressure of ≈ 0.6 GPa. $T_C$ has been reported to increase above this pressure reaching ≈ 69 K in $p = 2.08$ GPa. Some theoretical papers predicted a considerable increase of $T_C$ of the $VI_3$ monolayer under a tensile strain or having iodine deficiency[16, 17, 27].

Herein we report on the evolution of magnetism in $VI_3$ in external pressures up to 7.3 GPa which we found by detailed measurements of ac magnetic susceptibility (real part $\chi'$ and imaginary part $\chi''$) and low-field magnetization depending on temperature ($T$). We confirmed the existence of the two magnetic phase transitions reported in the literature[5, 6, 8, 10, 11]: a) a PM $\leftrightarrow$ FM1 transition at $T_C$ = 49.5 K, b) a transition between two ferromagnetic states FM1 and FM2 at $T_{FM}$ = 26 K. The temperatures $T_C$ and $T_{FM}$ in our nomenclature correspond to $T_{FM1}$ and $T_{FM2}$, respectively, introduced by Gati et al.[8] The $T_{FM}$-transition is most likely accompanied by the monoclinic $\leftrightarrow$ triclinic structural transition.[9]

Moreover, we observed visible peaks at temperatures $T_1$ = 54.5 K and $T_2$ = 53 K, on $\chi'(T)$, $\chi''(T)$, and $\partial M/\partial T$ vs. $T$ curves that allow understanding of the multistep anomaly observed in $M(T)$ dependence near $T_C$.[5] Considering results of the analysis of exchange interactions and theoretical DFT calculations of magnetism in $VI_3$[16, 27] we suggest that these anomalies reflect the onset of ferromagnetism in two types of surface layers of the $VI_3$ single crystal which are suffering from iodine deficiency or some lattice defects mimicking intralayer tensile strain. In this scenario, the intrinsic bulk ferromagnetism in $VI_3$ exists only at temperatures below $T_C$.

The characteristic temperatures $T_1$, $T_2$, and $T_C$ were found nearly intact by $p$ < 0.6 GPa but $T_{FM}$ increases dramatically with increasing pressure and the FM2 expands in the $T$-$p$ phase space on account of FM1. The FM1 phase is terminated at a triple point at ≈ 0.85 GPa. At higher pressures, only one ferromagnetic phase exists below $T_C$, which further increases with increasing $p$ and eventually reaches a high value of 99 K (double value of the ambient-pressure $T_C$) at 7.3 GPa. On the other hand, the high-field magnetization measured at 2 K dramatically decreases with increasing pressure up to 2.5 GPa. To support the interpretation of observed results selected first-principles calculations were also performed.

In this work, we tried to understand the origin of the newly observed strikingly complex evolution of magnetism in a two-dimensional material, accompanied by crystal structure changes with varying temperatures and high pressure. According to our data, this intriguing change arises from a complex hierarchy of magnetic and magnetoelastic interactions with pressure-induced details of the crystal structure and consequent electronic-structure changes.

## EXPERIMENTAL DETAILS

The $VI_3$ single crystals were grown by chemical vapor transport method according to the procedure described by Son et al.[5] The samples had a shape typical for vdW materials, thin plates with some hexagonal-like edges and the $c$-axis perpendicular to the plates at room temperature. Their typical lateral dimensions were up to several millimeters.

Magnetization and AC susceptibility data at ambient pressure were obtained utilizing MPMS XL 7T (*Quantum Design Inc.*) using a crystal oriented with the $c^*$-axis (≡ direction perpendicular to the $ab$ plane in the whole temperature range in contrast to the crystallographic $c$-axis) either parallel or perpendicular to the applied magnetic field. The magnetization behavior under hydrostatic pressures was measured with the same MPMS7T using two different pressure cells: a) for pressures up to 1 GPa the "piston-cylinder" pressure cell made from non-magnetic Cu/Be-alloy[37] with a Daphne 7474 pressure medium[38] and a superconducting lead manometer, determining the pressures directly at low temperatures, b) for high pressures up to 7.3 GPa, the "turnbuckle"-type diamond anvil cell (DAC)[39]. The cell is built from made-to-order ultrapure Cu/Be alloy, allowing sensitive measurements with an extremely small sample in DAC. Daphne 7575 pressure medium[40] was used, and pressure has been determined at room temperature by a ruby manometer. Due to geometrical constraints of the DAC cell the sample was oriented with the $c^*$-axis aligned to the applied magnetic field in the latter type of experiment.

The hydrostatic pressure influence of the AC susceptibility was measured using a set of homemade miniature coils with Ø of ≈ 1.5 mm and length of ≈ 6 mm (a primary coil for ac field generation has ~200 turns, secondary one for AC magnetic susceptibility detection having 50 and 50 turns wound in an opposite sense to compensate for the signal of the sample surroundings, see Fig. S8). The coils were connected to a Stanford Research Systems lock-in amplifier, model SR 850A used for AC signal processing while measuring both real ($\chi'$) and imaginary ($\chi''$) parts. The coils set fits inside double-layered CuBe/NiCrAl piston-cylinder pressure cell[41, 42] used to generate pressures up to ~ 3.5 GPa (room-temperature value). We used the Daphne oil 7373[43] as the pressure transmitting medium with a manganin manometer to determine the cell's pressure at room temperature. After cooling down to the base temperature, we calibrated the pressure again using our cell's pressure calibration table. The AC susceptibility measurement was accomplished by a Closed Cycle Refrigerator (*Janis Research Company, LLC/Sumitomo Heavy Industries, Ltd.*). The ambient-pressure specific-heat data were collected in the temperature range 2-300 K in magnetic fields up to 14 T applied along the *c**-axis using PPMS-14 (*Quantum Design, Inc.*). The angular dependence of magnetization was measured at 2 K using an MPMS sample rotator.

## CALCULATION DETAILS

Density functional theory (DFT) calculations were based on the full-potential linear augmented plane wave (FP-LAPW) method, as implemented in the ELK code[44]. The spin-orbit coupling has been included within the second variational approach. Generalized gradient approximation (GGA) has been employed. Since the system is known to be a Mott insulator[5, 7], Hubbard $U$ has been added to describe electron correlations within the fully localized limit double-counting treatment[45]. For $U = 3.8$ eV, this leads to predicting a bandgap with a size of around 0.6 eV. We covered the full Brillouin zone by sampling about 1000 $k$-points. We have assumed the rhombohedral structure of $BiI_3$ with experimental lattice parameters[7] $a = 0.6835$ nm and $c_0 = 0.6565$ nm (equilibrium interlayer distance).

To study the effect of pressure by first-principles calculations, we utilize the high anisotropy of compressibility in the vdW materials. And so up to some threshold, the compression takes place mainly perpendicular to the basal plane. We have assumed that the pressure leads to compression of the van der Waals gap within the studied range, while distances between V and I inside a layer remain unchanged.

## RESULTS AND DISCUSSION

Fig. 1 and 2 show how the magnetic phase transitions $VI_3$ are manifested in the temperature dependencies of specific heat, magnetization in a low static magnetic field, and AC susceptibility. The specific-heat measurement is one of the methods allowing to determine the critical temperature of magnetic ordering in a bulk material in the absence of a magnetic field. The lambda anomaly manifests a second-order magnetic phase transition of $VI_3$ at $T_C = 49.5$ K, as shown in Fig. 1(a). The evolution of the $C_p/T$ vs. $T$ plot with increasing magnetic fields displayed in Fig. S1 in Supplementary materials[46] is characteristic of ferromagnetic order. The value of $T_C$ determined here is reasonable with results published before [5-7, 10, 11].

The ferromagnetic transition at $T_C$ appears in the $\chi'(T)$ and $\chi''(T)$ curves as a high and sharp peak just at $T_C$, and as an inflection point of the steep increase of the low-field $M(T)$ dependences both in the field-cooled (FC) and zero-field-cooled (ZFC) mode upon cooling. The ferromagnetic ordering is accompanied by magnetostriction that distorts a crystal lattice. It is indicated by the change of slope of the temperature dependence of the angular position of the diffraction peak at $T_C$[9], in particular, the monoclinic (2 6 21) reflection.

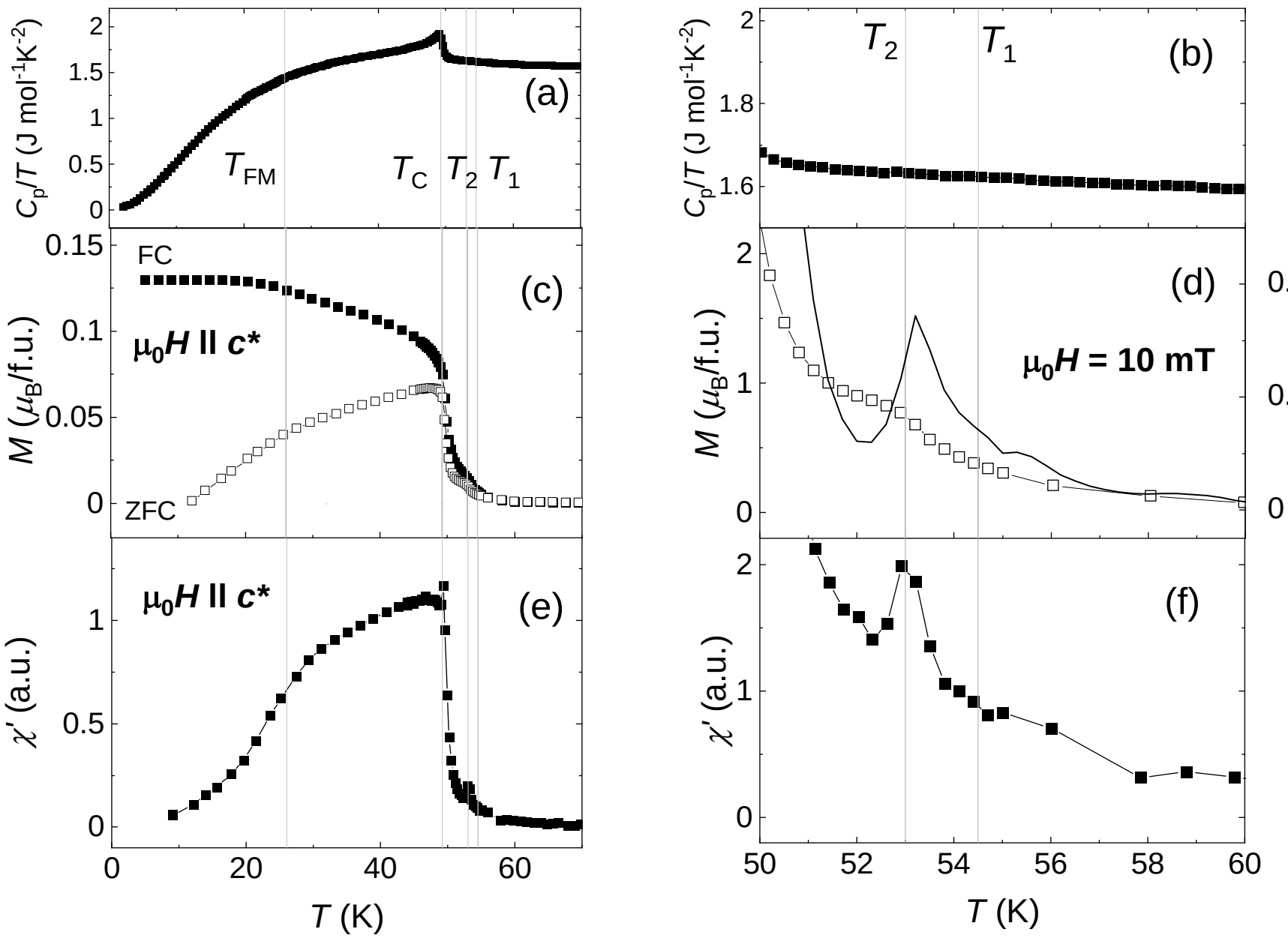

Fig. 1: The left-hand panels corresponding to the $Cp/T$ vs. $T$, $M(T)$, and $\chi'(T)$ plots show a broad temperature range including all the $T_1$-, $T_2$-, $T_C$-, and $T_{FM}$-transitions. The right-hand panels show their magnified images in a narrower temperature range above $T_C$ to see better the $T_1$- and $T_2$-related anomalies. Temperature dependences of specific heat ($Cp/T$ vs. $T$ plot) (a,b); FC and ZFC magnetization in a static magnetic field of 10 mT applied parallel to $c^*$-axis in the FC and ZFC mode, respectively (c,d) (the dashed line in (d) represents the - $\partial M/\partial T$ vs. $T$ plot); real component of the ac magnetic susceptibility $\chi'$ for the ac magnetic field applied along the $c^*$-axis (e,f). The error bars are smaller than the markers. The vertical lines represent the temperatures of the considered magnetic phase transitions.

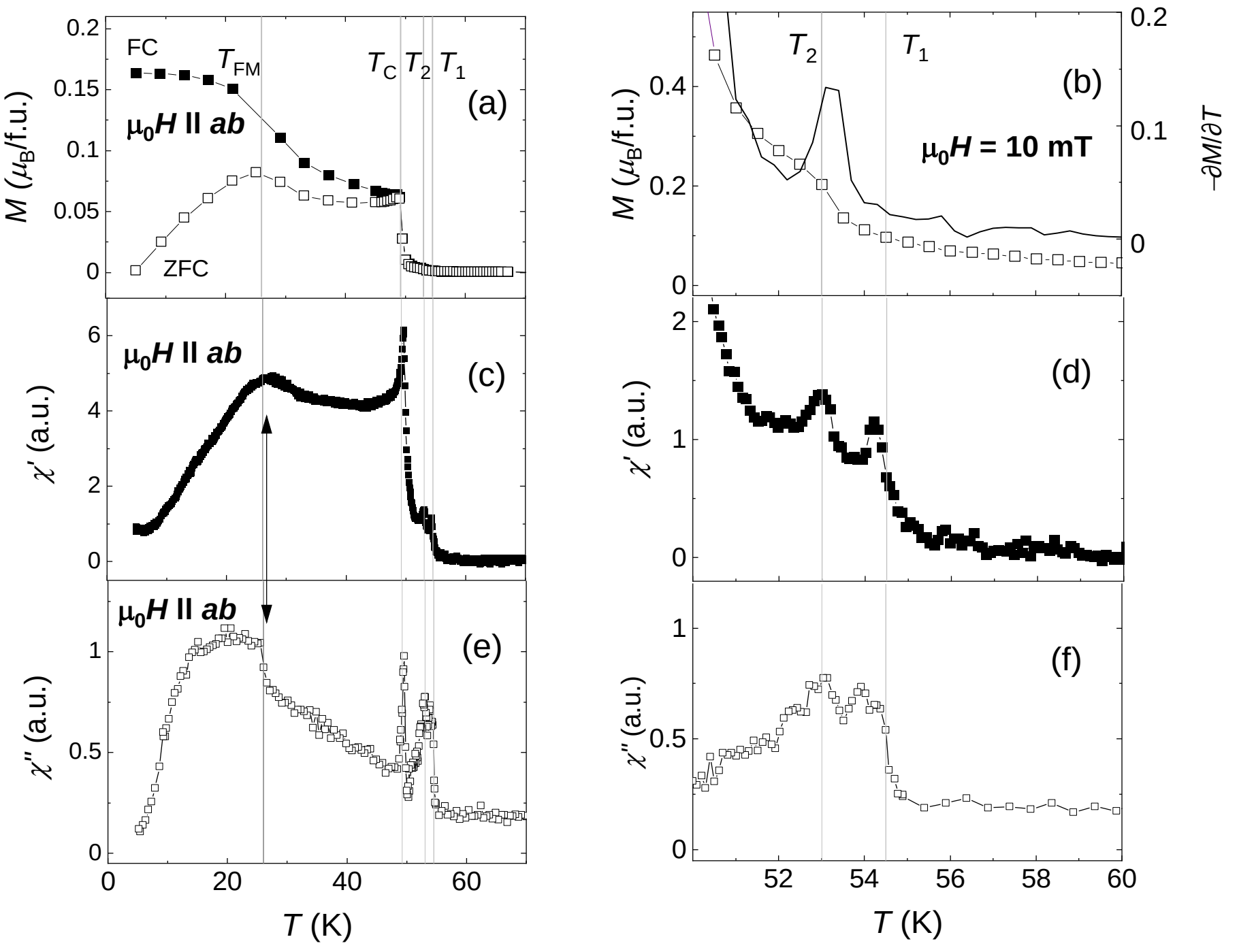

Fig. 2: The left-hand panels corresponding to the $M(T)$, $\chi'(T)$, $\chi''(T)$ plots show a broad temperature range including all the $T_1$-, $T_2$-, $T_C$-, and $T_{FM}$-transitions. The right-hand panels show their magnified images in a narrower temperature range above $T_C$ to see better the $T_1$- and $T_2$-related anomalies. The FC and ZFC

temperature dependences of magnetization in a static magnetic field of 10 mT applied perpendicular to the *c**-axis (a,b) (the dashed line in (b) represents the - $\partial M/\partial T$ vs. $T$ plot); real component of the ac magnetic susceptibility $\chi'$ for the ac magnetic field applied perpendicular to the the *c**-axis (c,d) and the corresponding imaginary component $\chi''(T)$. The error bars are smaller than the markers. The vertical lines represent the temperatures of the considered magnetic phase transitions.

Closer inspection of Figs. 1 and 2 reveals small peaks on the $\chi'(T)$, $\chi''(T)$, and $\partial M/\partial T$ vs. $T$ curves at temperatures above $T_C$, namely at $T_1$ = 54.5 K and $T_2$ = 53 K. Analogous results were reported by Liu et al.[11] and Gati et al.[8] Measuring different samples of the $VI_3$ crystals we have found that the size of these anomalies is strongly sample dependent (in some cases only the peak at $T_2$ is visible) but the differences $T_1 - T_C$ = 5 K and $T_2 - T_C$ = 3.5 K, respectively, seem to be sample independent. The lack of any response around $T_1$ and $T_2$ on the $C_p/T$ vs. $T$ plots corroborate the idea that the $T_1$- and $T_2$-transitions are in fact not intrinsic phenomena in the bulk of the $VI_3$ crystal. $\partial M/\partial T$ vs. $T$ plots, $\chi'(T)$, and especially $\chi''(T)$ anomalies at $T_1$ and $T_2$ provide strong indications of ferromagnetic transitions in parts of samples.

Theoretical studies within the density functional theory revealed that the hierarchy of exchange interactions in a $VI_3$ monolayer could be influenced by applying tensile strain[16] or producing iodine deficiency[27] to increase Curie temperature. Considering these results and having in mind that the $VI_3$ surface rapidly degrades and finally decomposes when exposed to oxygen and moisture, we suggest that two types of surface iodine-poor layers may become ferromagnetic at temperatures ($T_1$ and $T_2$, respectively) higher than $T_C$ of the $VI_3$ bulk crystal.

The $C_p/T$ vs. $T$ plot at temperatures below 40 K is very smooth and monotonous, showing no trace of the $T_{FM}$-transition, not even the weak cusp at 36 K observed with a sample in pressure cell[8]. The characteristic temperature of structural transition was reported at 32 K[9] at which we also cannot find clear indications of a magnetic phase transition in the AC susceptibility data.

On the other hand, the maximum in the $\chi'(T)$ and low-field $M(T)$ curves and particularly the discontinuous step in $\chi''(T)$ dependence at $T_{FM}$ = 26 K in fields perpendicular to the *c**-axis point to a sudden change of ferromagnetic phase, probably a transition between the two ferromagnetic phases (FM1 existing at temperatures between $T_C$ and $T_{FM}$ and the ground state phase, FM2, below $T_{FM}$). The evolution of these anomalies with applied hydrostatic pressure and comparison with the magnetic phase diagram presented in Ref. 7 discussed below corroborate this idea.

In Fig. 3, we can see how the $\chi'(T)$ and $\chi''(T)$ dependences are modified by applying hydrostatic pressure in the $VI_3$ crystals. The positions of features related to $T_1$, $T_2$, and $T_C$ appear not to be affected by pressures up to $\approx$ 0.6 GPa. Then these characteristic temperatures slightly increase with increasing $p$ up to $\approx$ 0.8 GPa. $T_{FM}$, on the contrary, increases rapidly with increasing pressure already in the lowest pressures, and for $p \approx$ 0.8 GPa, it reaches $T_C$. For pressures higher than 0.8 GPa, we can see only one anomaly on the $\chi'(T)$ and $\chi''(T)$ curves, respectively.

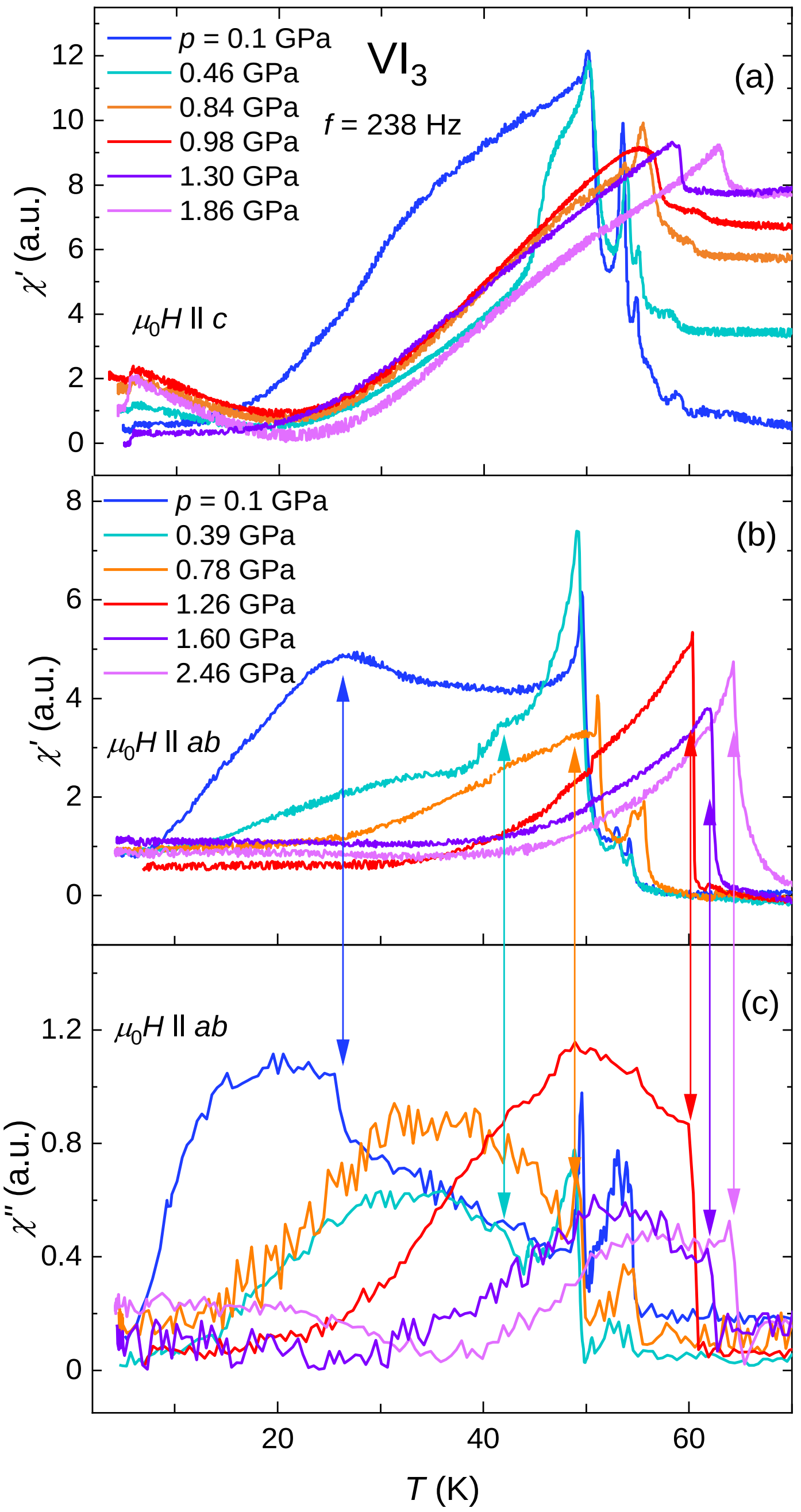

Fig. 3: Selected $\chi'(T)$ dependences for the ac magnetic field applied parallel to the $c^*$-axis (a) and $\chi'(T)$ and $\chi''(T)$ curves measured in the ac field in the *ab*-plane (b,c, respectively) on the $VI_3$ crystals exposed to various hydrostatic pressures. The double-sided arrows indicate the $\chi'(T)$ and $\chi''(T)$ anomalies corresponding to the $T_{FM}$-transitions observed at particular pressures.

The 1-mT $M(T)$ dependences measured in pressures up to 7.3 GPa using the DAC are displayed in Fig. 4. The $T_C$ value was determined as the temperature of the inflection point of the steep $M(T)$ increase with decreasing temperature. The $M(T)$ curve measured in 0.4 GPa exhibits a visible sidestep at ≈ 53 K, which can be understood as a mark of the $T_2$-transition. Consistently with $\chi'(T)$ data, in which the $T_2$-transition anomaly does not show up at pressures above 1 GPa, the $M(T)$ curve is smooth at $T > T_C$. The three $M(T)$ curves in pressures 0.78 – 3.5 GPa are incredibly sharp in the vicinity of $T_C$ (within 1 K), which may signalize the first-order transition as also suggested by Gati et al.[8]

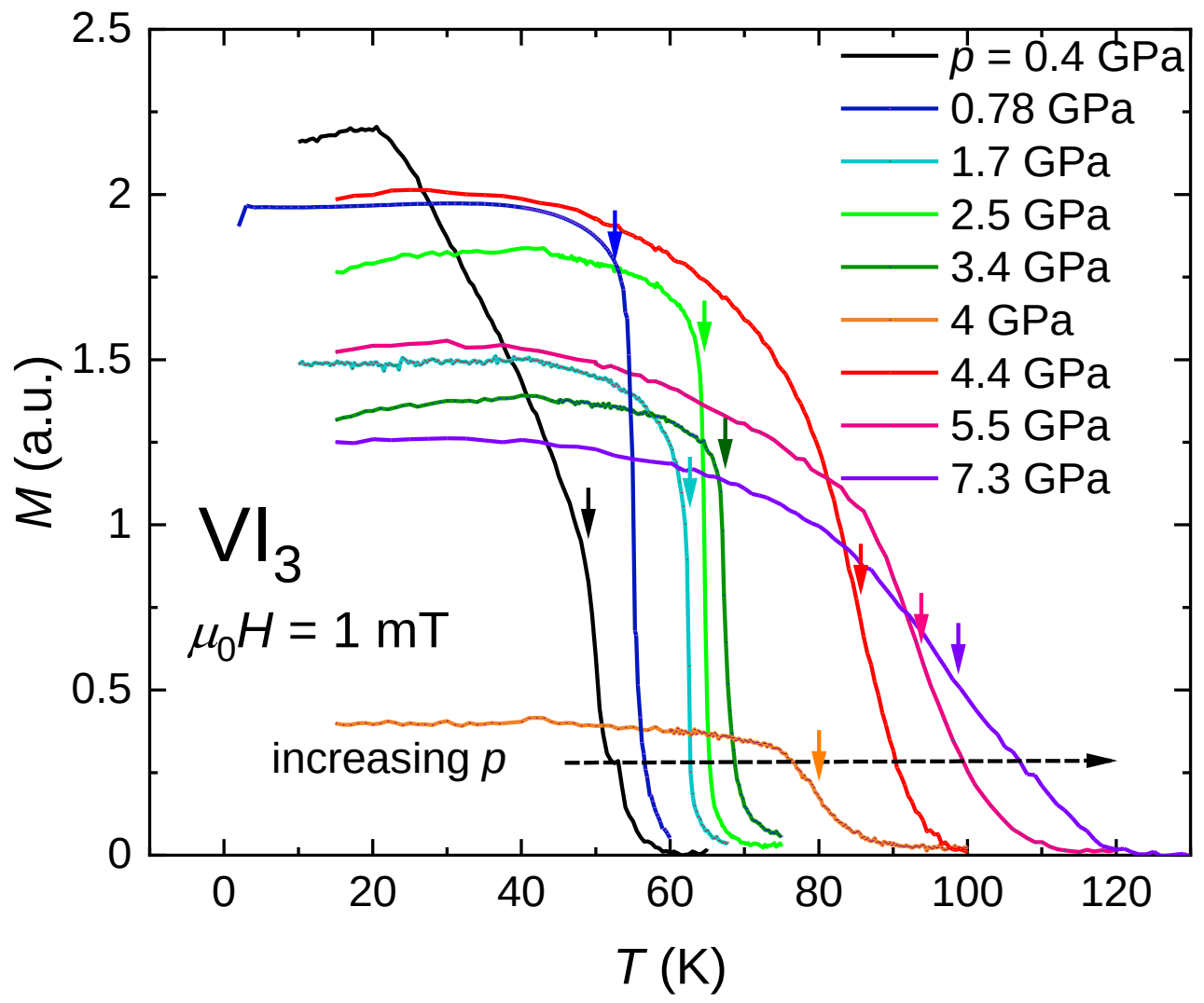

Fig. 4: Temperature dependences of magnetization in a static magnetic field of 1 mT applied parallel to the $c$*-axis of a $VI_3$ single crystal exposed to high pressures generated by DAC. The point at 0.78 GPa originates from the piston-cylinder cell (see Fig S4 [46]). The arrows mark the positions at which the transition temperatures $T_C$ were determined.

The pressure dependences of characteristic temperatures $T_1$, $T_2$, $T_{C,}$ and $T_{FM}$ obtained from data analysis are presented in the schematic $T$-$p$ phase diagram in Fig. 5. The $T_1(p)$ and $T_2(p)$ lines simultaneously slightly increase the pressure up to ≈ 1 GPa. Since the $T_1$- and $T_2$-transitions are not intrinsic magnetic phase transitions of bulk $VI_3$, they will not be further discussed in the framework of the $T$-$p$, the magnetic phase diagram of this compound.

The intrinsic magnetic phases and phase lines in the $T$-$p$ phase space for $p \leq 2.08$ GPa derived from data of our AC susceptibility and low-field magnetization data vs. temperature are in good agreement with the results of Gati et al.[8] who discovered that $VI_3$ appears in two ferromagnetic phases. The transition between two ferromagnetic phases occurring at a critical temperature is designated in our nomenclature as $T_{FM}$. When cooled at ambient pressure, $VI_3$ first undergoes the transition from PM to FM1 state at $T_C$ and then at $T_{FM}$ (< 40 K) a transition between two ferromagnetic phases. Our specific-heat, ac-susceptibility, and low-field magnetization data obtained at ambient pressure point to $T_C$ = 49.5 K and $T_{FM}$ = 26 K (except for specific heat). The Curie temperature does not change significantly below 0.6 GPa, but $T_{FM}$ increases fast with increasing the applied pressure. Consequently, the FM2 phase expands in the $T$-$p$ phase space on account of the FM1 one. $T_C(p)$ and $T_{FM}(p)$ phase lines finally meet at the triple point in pressure $p_{TP}$ near to 0.8 GPa, where the FM1 phase terminates. Although there has been speculation about this point's tricritical nature in the $T$-$p$ phase space[8], we judge that more pressure points would need to be measured to answer this question. In pressures $p > p_{TP,}$ only the phase FM2 exists with the critical temperature labeled as $T_C$. The temperature $T_C$ then further increases with increasing pressure. The evolution of bulk magnetic phases and related phase transitions with pressure up to 2.08 GPa observed in our study is in good agreement with results published by Gati et al.[8] They also showed that the critical temperature of the monoclinic ↔ trigonal structural transition $T_s$ rapidly decreases with applied hydrostatic pressure and the $T_s(p)$ and $T_C(p)$ curves merge at $p_m \approx 1.35$ GPa. That probably means that the ferromagnetic transition at $T_C$ is accompanied by the trigonal ↔ monoclinic or trigonal ↔ triclinic structural transition for $p > p_m$. In contrast, at lower pressures, the monoclinic structure remains preserved at temperatures above $T_C$ up to $T_s$.

At higher pressures, the $T_C(p)$ almost saturates around 2.5 GPa but resumes considerable growth in pressures above 3.4 GPa. Finally, at the maximum pressure of our experiment, 7.3 GPa, $T_C$ reaches 99 K. This is the double value of the ambient-pressure $T_C$ value. The huge pressure-induced increase in Curie temperature of $VI_3$ belongs to the most important results of this study.

The accelerated increase of $T_C$ with increasing pressure in the 3.4 - 4 GPa pressure range can be due to the non-hydrostaticity of the pressure transmitting medium (Daphne 7575) in DAC in the high-pressure range. In this case, partially uniaxial stress along the $c^*$-axis is expected. This result appears to be consistent with the calculated effect of the $c^*$-axis compression on the exchange integral shown in Fig. 6.

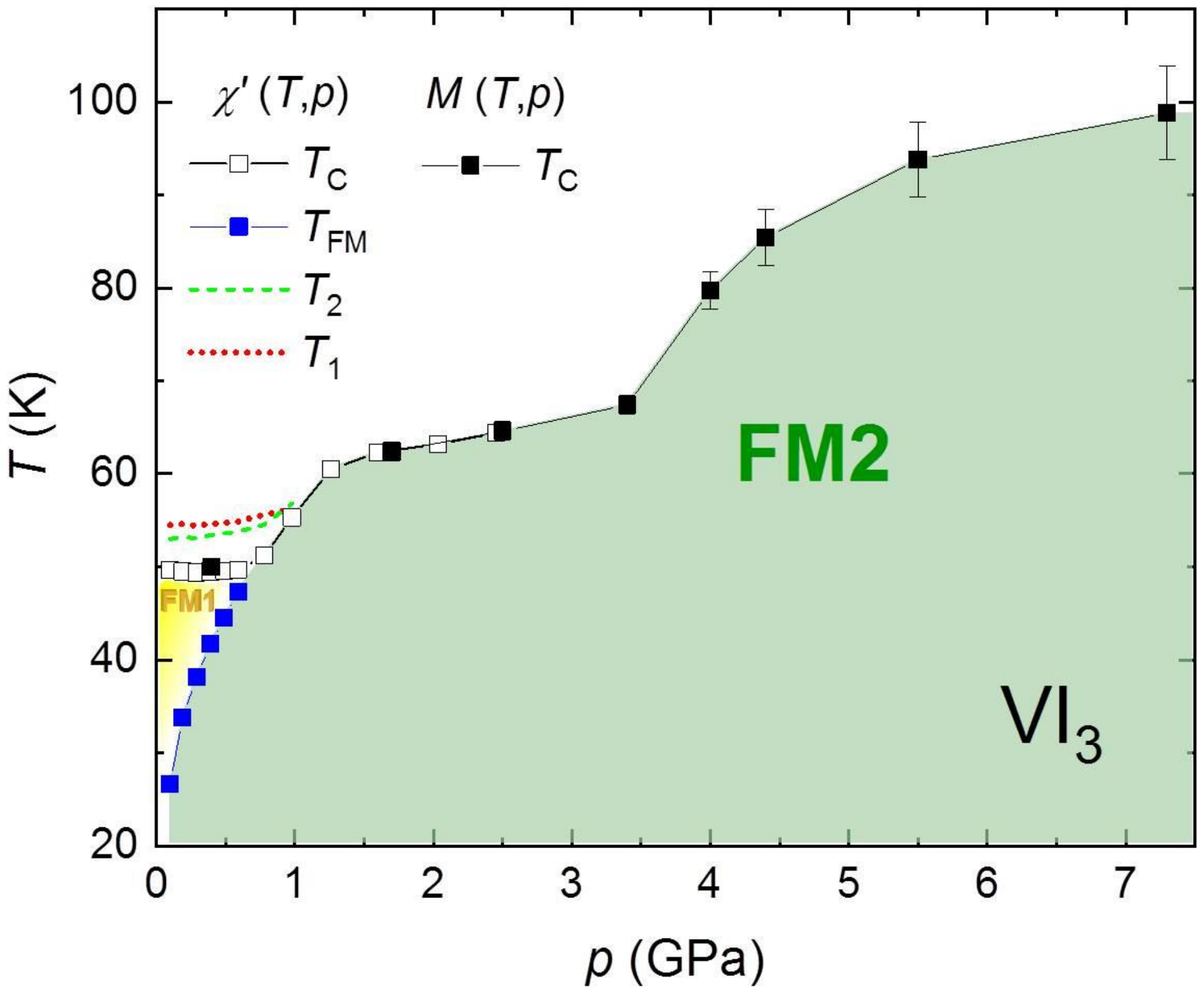

Fig. 5: Temperature-pressure *(T-p)* magnetic phase diagram of $VI_3$ derived from results of the ac and DC magnetic susceptibility magnetization measurements measured with a sample in a piston-cylinder and DAC pressure cell, respectively. The meaning of the critical temperatures $T_1$, $T_2$, $T_C$, and $T_{FM}$, as well as the phases FM1 and FM2 is explained in the text. The comparison of data sets measured by DAC and two different piston-cylinder cells, respectively, in the overlapping pressure interval (≤ 2.5 GPa) proves that the pressures accomplished by the two different types of cells are coinciding.

First-principles calculations have been employed to examine the effect of lattice compression on magnetic interactions, which control the Curie temperature. EFM characterizes intralayer exchange interactions - $E_{nAFM}$, which corresponds roughly to 6 J1 + 6 J3 in nearest-neighbor interactions between V atoms.[27] We found these interactions are strong with $E_{FM} - E_{nAFM}$ = 71 meV, in agreement with another calculation where these values correspond to $T_C$ ~ 240K in layers[27]. These robust interactions originate from the nearly 90° FM superexchange.[47] And it is not affected by the vdW gap's compression up to a specific limit. The interlayer interactions are described by $J_z = E_{FM} - E_{zAFM}$, where zAFM corresponds to antiferromagnetically ordered layers. This coupling is much weaker, with $J_z$ = 5meV for the equilibrium structure, similar to layered Cr compounds[48]. It is thus reasonable to expect that in bulk, the $T_C$ will be affected mainly by the interlayer interaction $J_z$. The value of this interaction as a function of interlayer distance $c$ is plotted in Fig. 6. For small compressions, it remains almost constant, while it grows approximately linearly for larger ones, in a surprising agreement with the measured $T_C$. It shows that the vdW gap's compression leads to the observed two-phase increase of $T_C$, without the need for the suggested sudden onset of interplanar interactions.[5] There is also no need to consider pressure-induced modification of the structure of individual planes, which was predicted to lead to a $T_C$ increase too.[16, 17, 27] Although after passing a specific threshold pressure value, this may become important too.

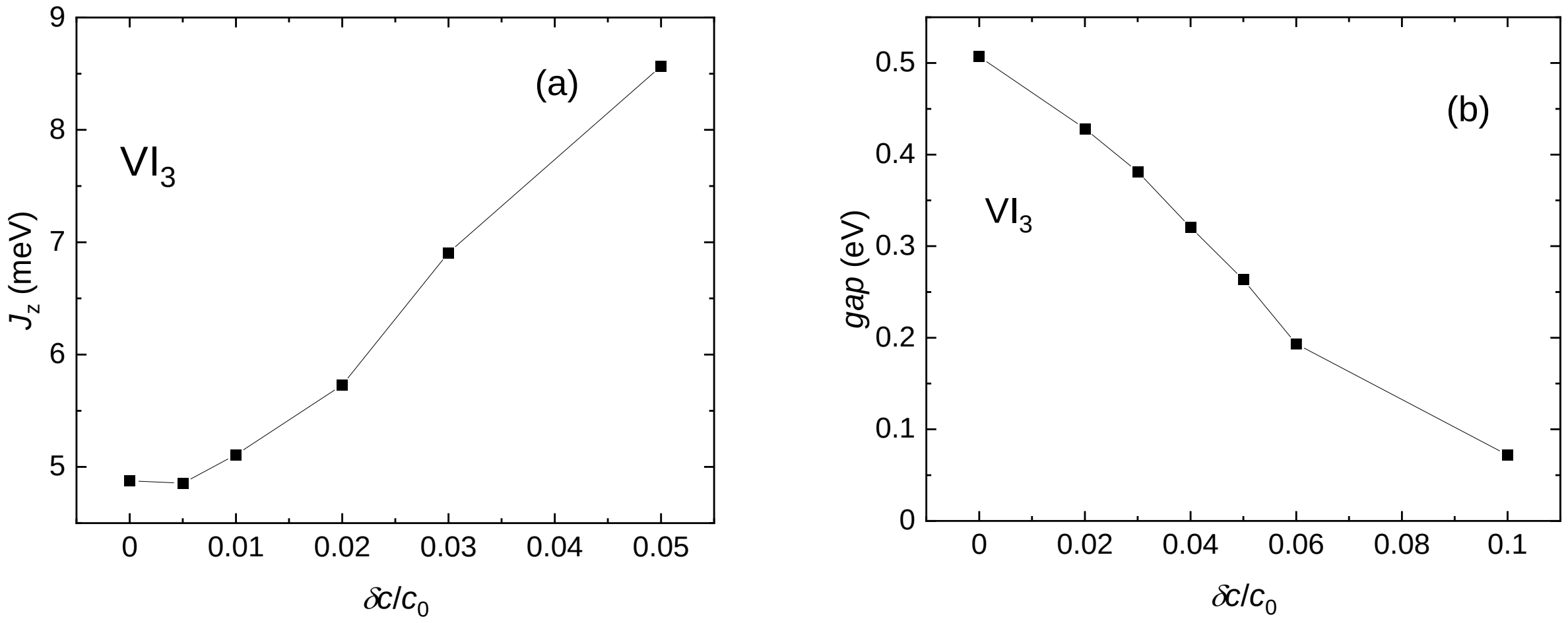

Fig. 6: Calculated dependence of the exchange coupling $J_z = E_{FM} - E_{zAFM}$ (a) and the energy gap (b) on the relative compression of the $VI_3$ single crystal along the *c**-axis.

The low-temperature magnetization data measured on $VI_3$ single crystals in magnetic fields parallel and perpendicular to the *ab*-plane, respectively, provide information on a magnetocrystalline anisotropy[5, 6, 10]. In contrary to the uniaxial anisotropy in $CrI_3$[29], the situation in $VI_3$ appears to be more complicated. In this case, the anisotropy is strong, and the easy magnetization direction is undoubtedly not perpendicular to the *ab*-plane. However, the *c**-axis magnetization in fields > 2 T is considerably larger than the corresponding values in perpendicular direction[5, 6, 10]. Kong et al.[6] pointed out the specific *M*(*H*) behavior can be understood in terms of V magnetic moments canted from the *c**-axis. We measured the angular dependence of the 5T-magnetization in the *c**-*a* plane at 2 K.

The $M(\phi)$ dependence shown in Fig. 7 has a minimum for $H \perp c^*$, a local minimum for $H // c^*$, and a maximum for the angle of canting $\phi = 40°$ from the *c**-axis. Similar results were reported by Yan et al.[10] In the simplest case of collinear V magnetic moments canted by 40° from the *c**-axis, one can consider that the $M // c^*$ and $M \perp c^*$ data represent their projections on the *c**-axis and the *ab*-plane. The difference between $VI_3$ and $CrI_3$ may originate from the fact that $CrI_3$ has a closed $t_{2g}$ shell, unlike $VI_3$.

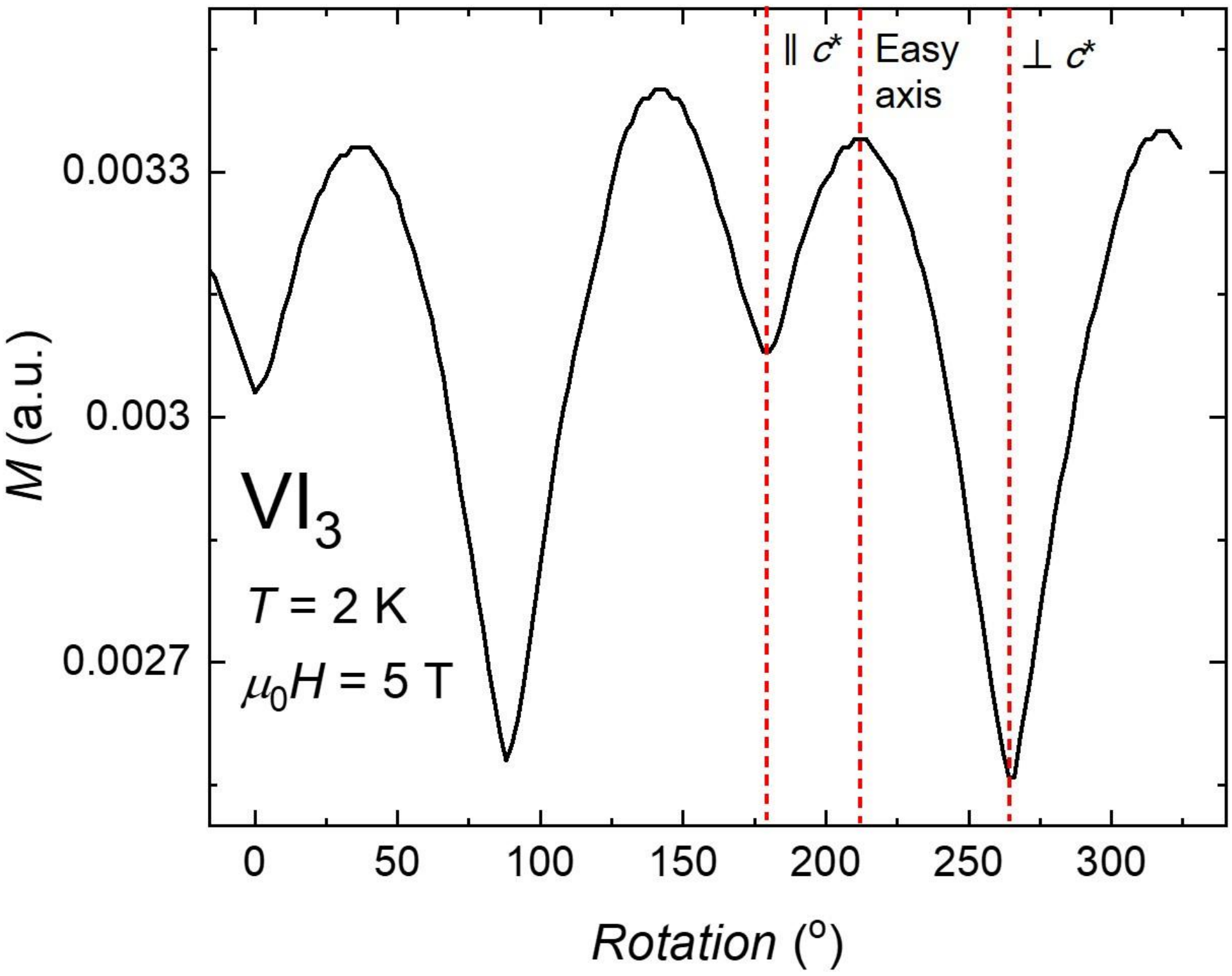

Fig. 7: Angular dependence of the ambient-pressure magnetization in the *c**-*a* plane of a $VI_3$ single crystal at 2 K.

The thermomagnetic curves and hysteresis loops of $VI_3$ single crystals were measured in various pressures in magnetic fields applied in two principal directions, i.e., parallel and perpendicular to the *c**-axis at selected temperatures. The $M(H)$ hysteresis loops in Fig. 8 show that in both cases, the high-field magnetization considerably decreases with increasing pressure already in relatively small pressures (<1 GPa), which indicates the pressure-induced reduction of the V moment (by ~ 14% between 0.1 and 0.74 GPa). This measurement was done with the 1 GPa piston-cylinder cell. The extended measurements to 2.5 GPa are shown in Fig. S7. There we can see that the *c**-axis magnetization in high fields decreases with increasing pressure by almost the same rate as in the lower-pressure measurements performed with the 1 GPa piston-cylinder cell. If this reduction in the measured magnetization is proportional to the change in the magnetic moment of the ion V, it would mean that the V moment decreases to half of its ambient-pressure value due to the application of $p$ = 2.5 GPa (see also Fig. 8c). Such an effect of pressure on the magnetic moment of Vanadium is challenging to imagine in insulators. The rate of moment reduction for field along the *c**-axis is higher than that in perpendicular direction. This may indicate an increasing angle of canting of moments from the *c**-axis with increasing pressure between 0 and 2.5 GPa.

The high-field magnetization decreases at a very similar rate in pressures up to 2.5 GPa. At higher pressures from 3.4 to 7.3 GPa, it almost does not change. It may be connected with the non-hydrostaticity as mentioned above of pressure due to the frozen medium in DAC in this pressure range. In this case, preferentially uniaxial stress along the *c**-axis is expected, whereas the in-ab-plane compression increment may become negligible. If compression in the in-*ab*-plane has a dominant effect on the V magnetic moment, the uniform magnetization can be understood at the highest pressures.

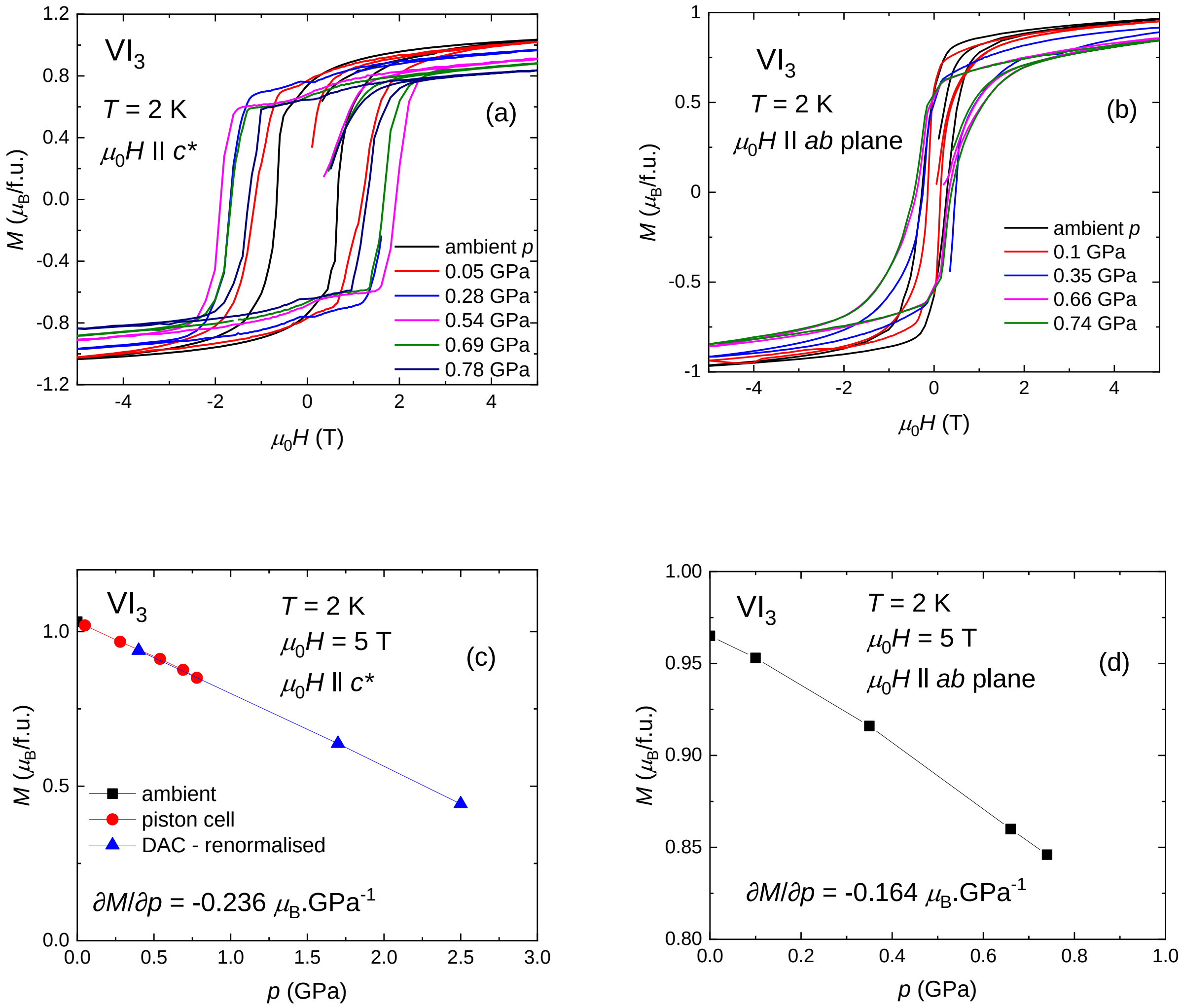

Fig. 8: Magnetization hysteresis loops of a $VI_3$ crystal exposed to several hydrostatic pressures measured at 2 K in magnetic fields applied parallel (a) and perpendicular (b) to the *c**-axis. Pressure dependence of the magnetic moment of $VI_3$ measured at 2 K in the magnetic field of 5 T applied parallel (c) and perpendicular (d) to the *c**-axis.

An explanation of this result would be possible in case of pressure-induced delocalization of the 3*d* electrons and some kind of metallization of $VI_3$ slowly progressing already from low pressures, which is, however, highly unlikely. Our calculations show that the interlayer distance compression leads to a linear decrease of the bandgap (Fig. 6), which would finally close if the interlayer distance *c* would be reduced by $\delta c \sim 0.1\ c_0$, where $c_0$ is the equilibrium interlayer distance. Pressure-induced insulator-metal transition can thus be expected, but not at low pressures.

The discovery of a pressure-induced insulator-metal transition in a vdW compound $FePS_3$ accompanied by a transition from ambient-pressure antiferromagnetism (AFM) to high-pressure metallic ferromagnetism was recently published[35]. The thorough study by X-ray and neutron diffraction revealed a complex pressure AFM → FM transformation (evolution), composed of several stages extended to much lower pressures than the critical pressure of the Mott transition.[35, 36] Therefore, we feel that the puzzle of the pressure-induced evolution of V magnetic moments can be resolved only after a thorough study based on X-ray and neutron diffraction and possibly μSR spectroscopy studies.

## CONCLUSIONS

In this study we have observed four magnetic phase transitions at temperatures $T_1$ = 54.5 K, $T_2$ = 53 K, $T_C$ = 49.5 K, and $T_{FM}$ = 26 K, respectively. The upper two transitions are attributed to the onset of ferromagnetism in the crystal-surface layers caused by a deficiency of iodine or some specific lattice defects preceding the crystal surface's decomposition. The Curie temperature of the bulk ferromagnetism is seen in a λ-shape anomaly of the temperature dependence of specific heat and relevant experiments. The temperature $T_{FM}$ is a characteristic of a transition between two different ferromagnetic phases, which is accompanied by a structure transition from monoclinic to triclinic symmetry upon cooling. The first three transitions are only slightly affected by pressures up to 0.6 GPa, whereas the $T_{FM}$-transition moves fast to higher temperatures with increasing pressure. Only one magnetic (ferromagnetic) phase transition at 55.5 K is observed in pressures > 0.9 GPa. $T_C$ then further increases with increasing pressure. The $T_C(p)$ increase almost saturates around 2.5 GPa. A rapid increase of $T_C$ is restored in pressures above 3 GPa to reach 99 K at 7.3 GPa. In contrast to increasing $T_C$, the low-temperature bulk magnetization is dramatically reduced by pressures up to 2.5 GPa, indicating a possible pressure-induced reduction of vanadium magnetic moment. The results are discussed in recent publications including the theoretical studies that analyzed exchange interactions and suggested ways to increase the Curie temperature of $VI_3$.

A thorough study of crystal structure and microscopic aspects of magnetism by X-ray and neutron diffraction and possibly μSR spectroscopy studies of the $VI_3$ single crystals under pressure is needed to resolve the mechanism responsible for the results presented in this paper.

## ACKNOWLEDGMENTS

This work is part of the research program GACR 19-16389J which is financed by the Czech Science Foundation. Works at SNU were supported by the Leading Researchers Program of the National Research Foundation of Korea [NRF-2020R1C1C1013642] and by the Institute for Basic Science [IBS-R009-G1] of the Republic of Korea. Experiments were performed in MGML (https://mgml.eu/), which is supported within the program of Czech Research Infrastructures (project no. LM2018096).

**SUPPORTING MATERIALS**

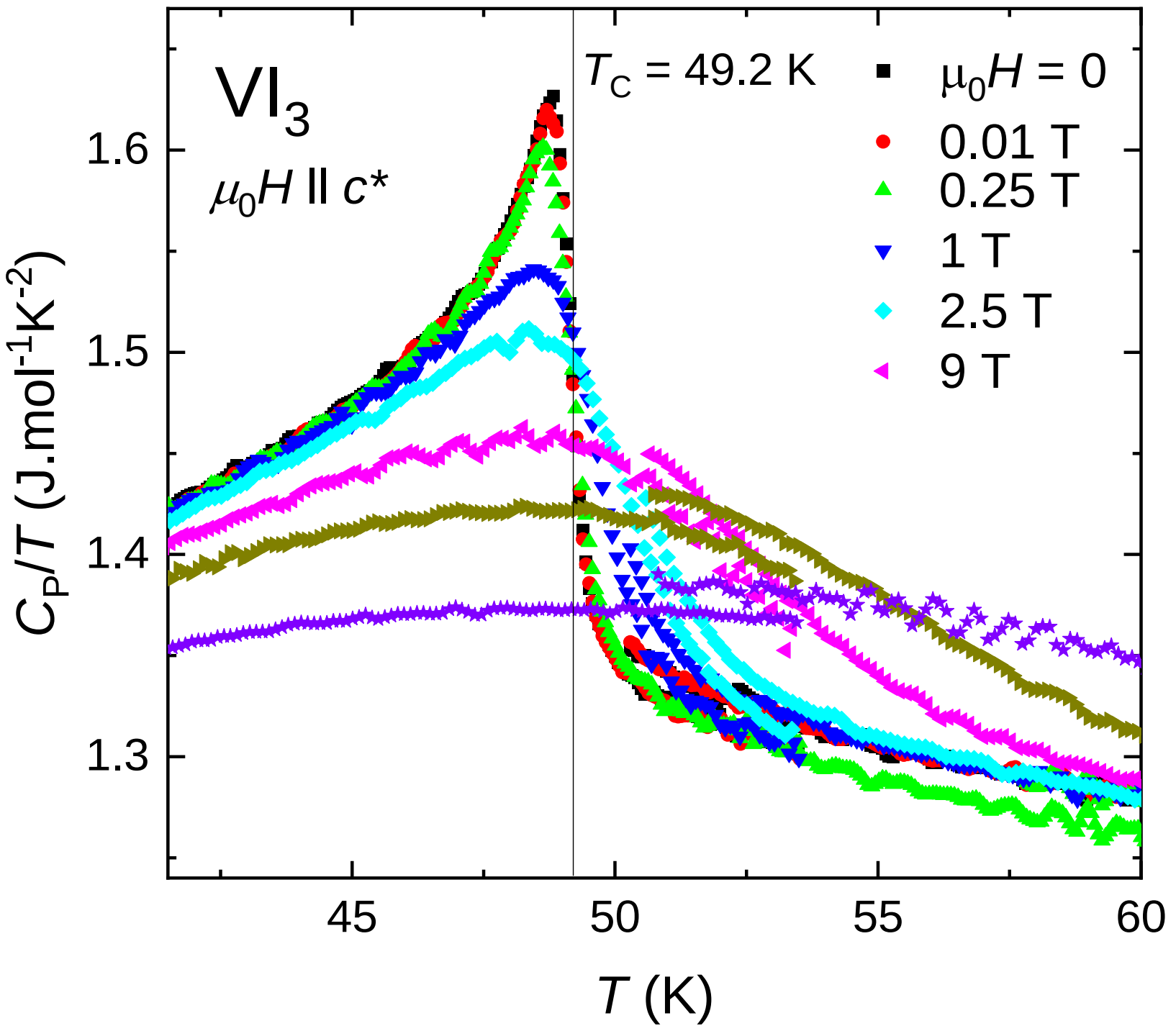

FIG. S1: The lambda-shape anomaly in the temperature dependence of the specific heat of $VI_3$ in zero magnetic fields reflecting the second-order phase transition at $T_C$ and its evolution with applied magnetic field documenting that it is a transition to the ferromagnetic state below $T_C$.

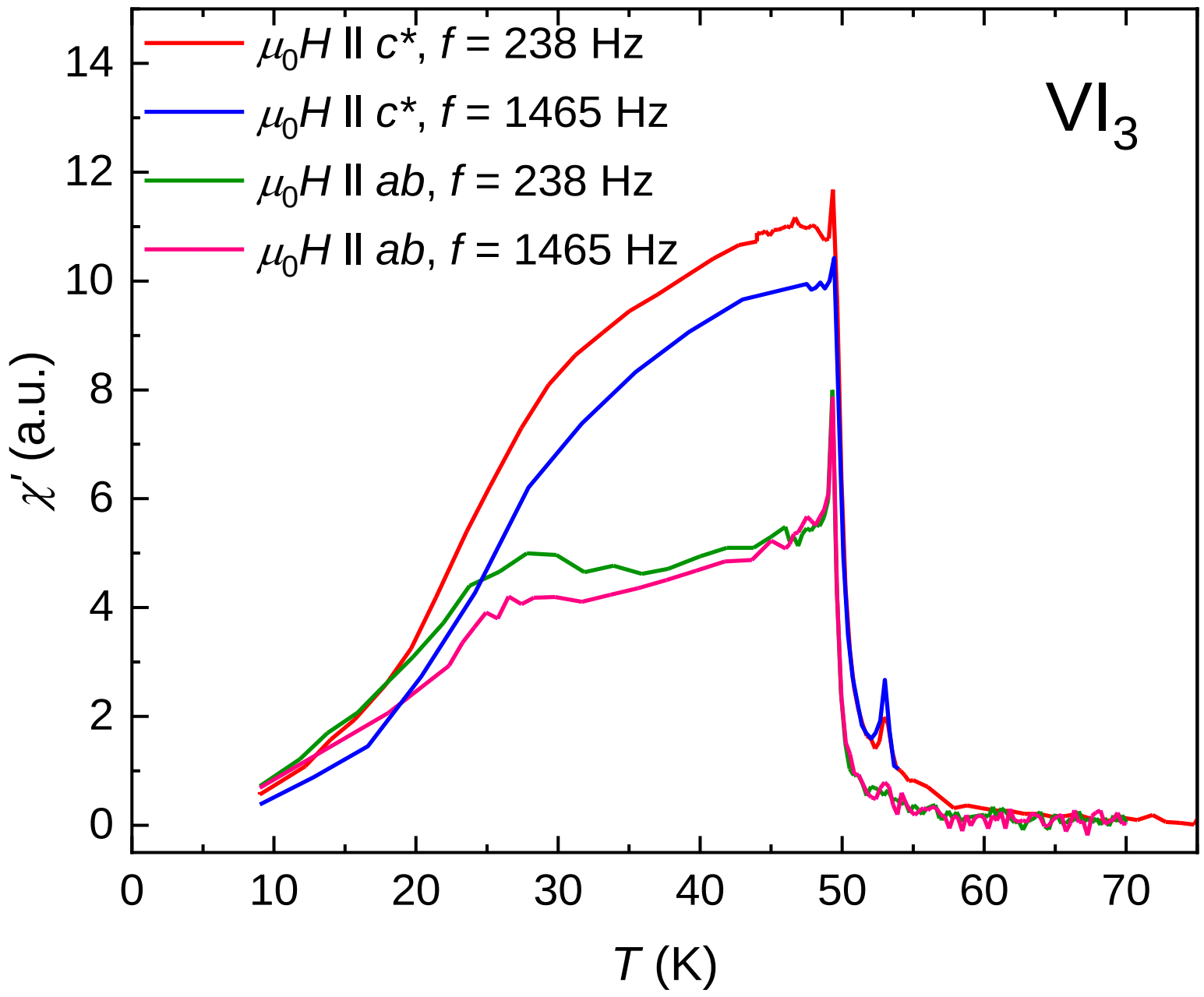

FIG. S2: $\chi'(T)$ dependences measured with the ac magnetic field applied parallel and perpendicular to the $c^*$-axis at two different frequencies.

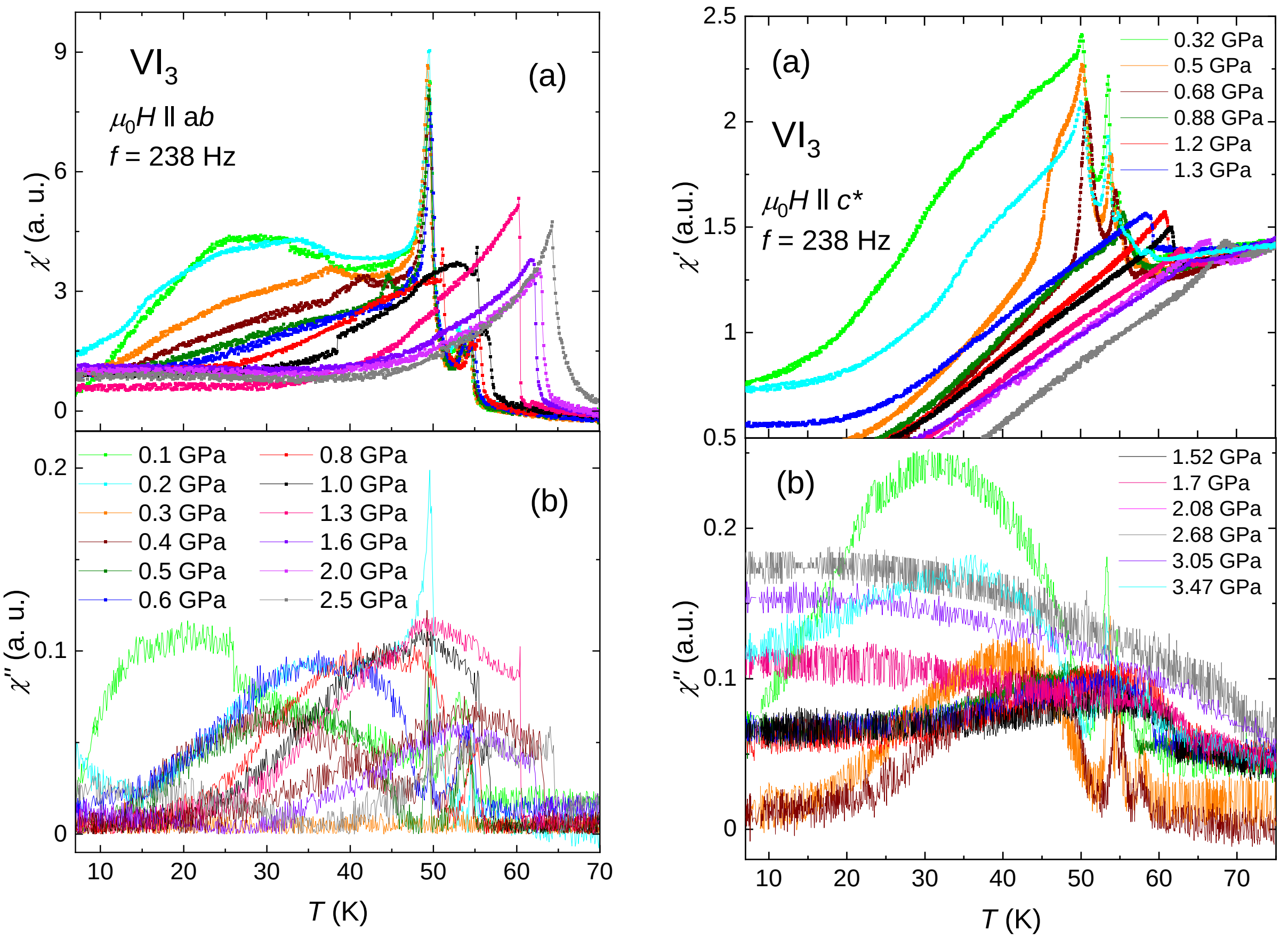

FIG. S3: All real (a) and corresponding imaginary (b) temperature dependences measured with the ac magnetic field ($f$ = 238 Hz) applied perpendicular (left panel) and parallel (right panel) to the $c$*-axis on the $VI_3$ single crystals exposed to various hydrostatic pressures.

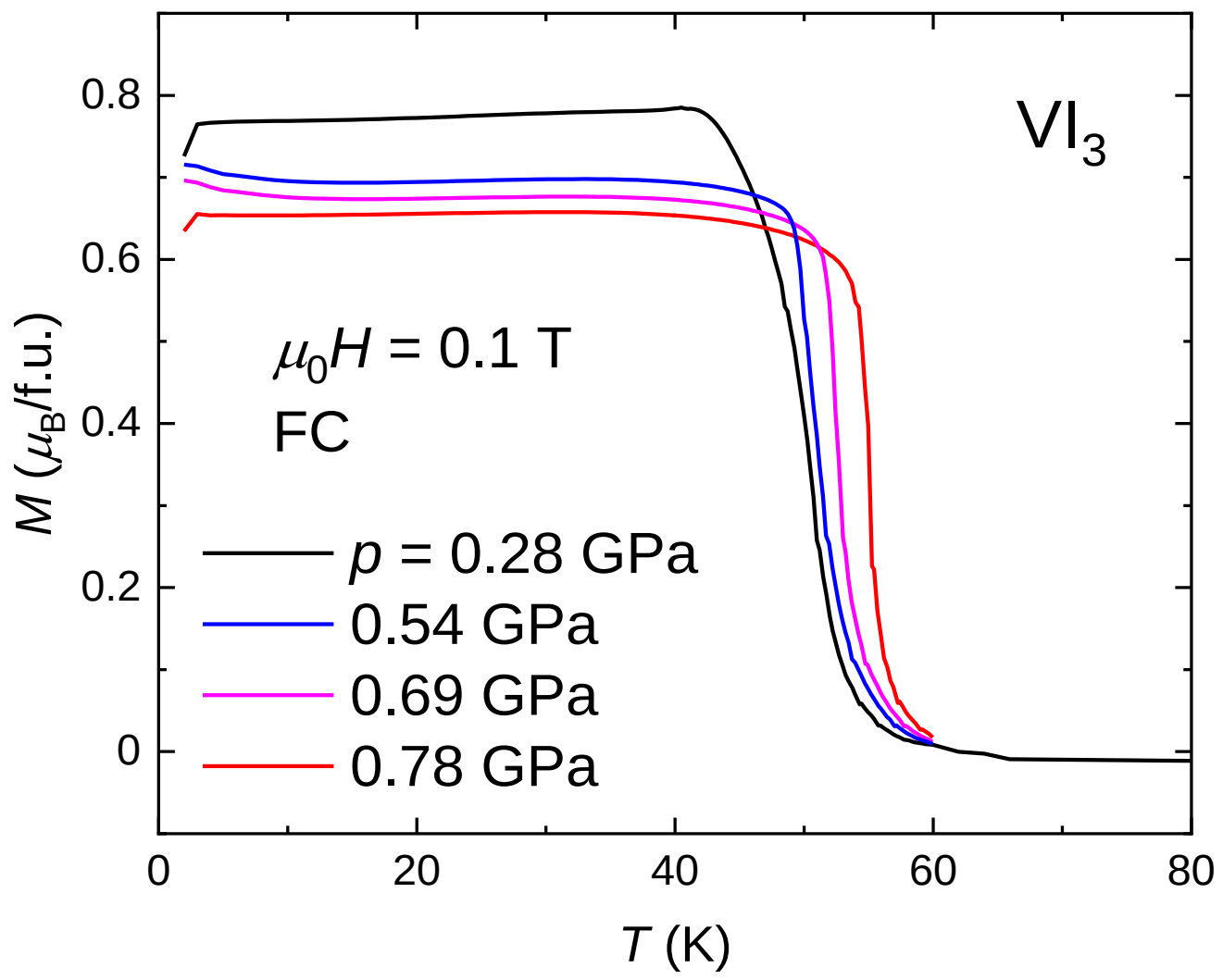

FIG. S4: Temperature dependence of magnetization in a static magnetic field of 100 mT in the field-cooled mode applied along the $c$*-axis of a $VI_3$ single crystal exposed to hydrostatic pressures generated by the piston-cylinder cell.

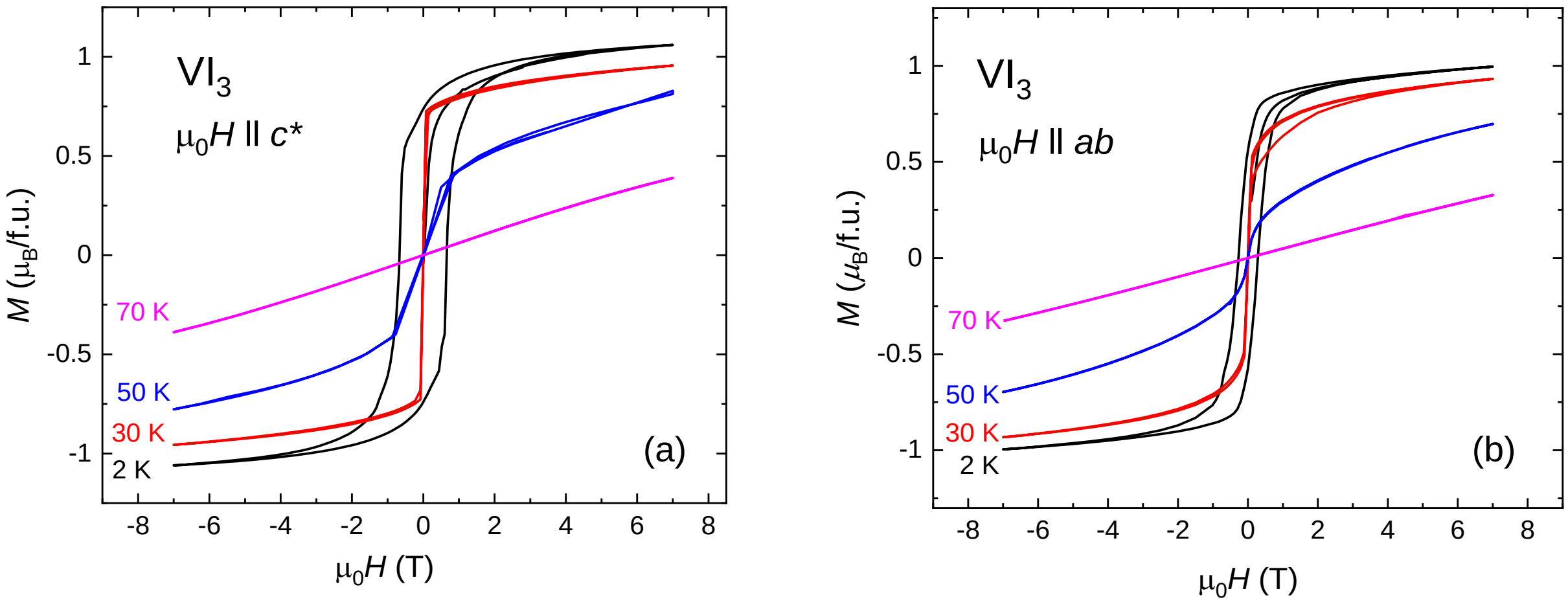

FIG. S5: Magnetization hysteresis loops of a $VI_3$ crystal measured in ambient pressure at 2 K in magnetic fields applied along (a) and perpendicular (b) to the *c**-axis.

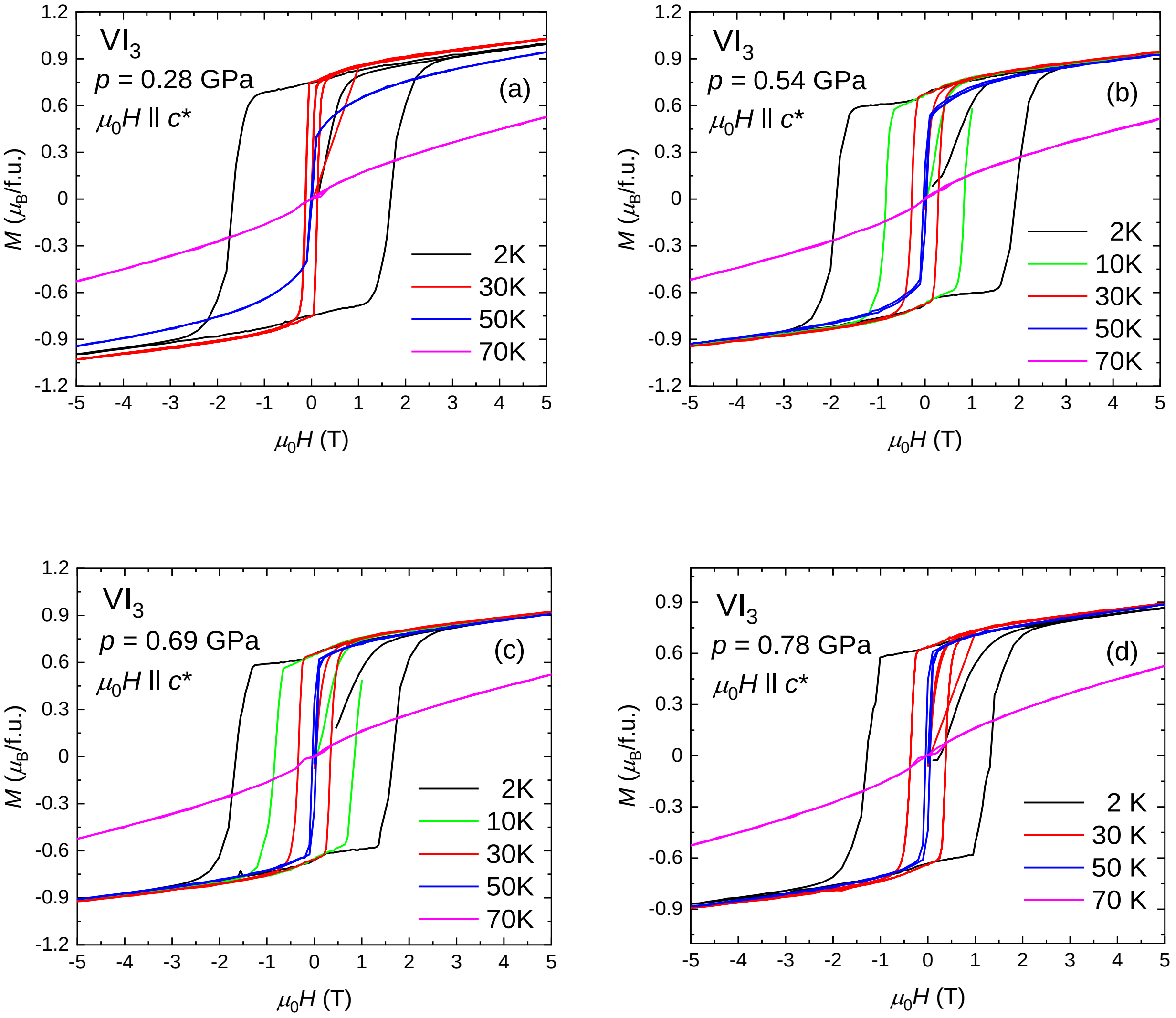

FIG. S6: Magnetization hysteresis loops of a $VI_3$ crystal measured at 2, 30, 50, and 70 K, respectively, in hydrostatic pressure of 0.28 GPa (a), 0.54 GPa (b), 0.69 GPa (c) and 0.78 GPa (d) in the magnetic field applied along the *c**-axis.

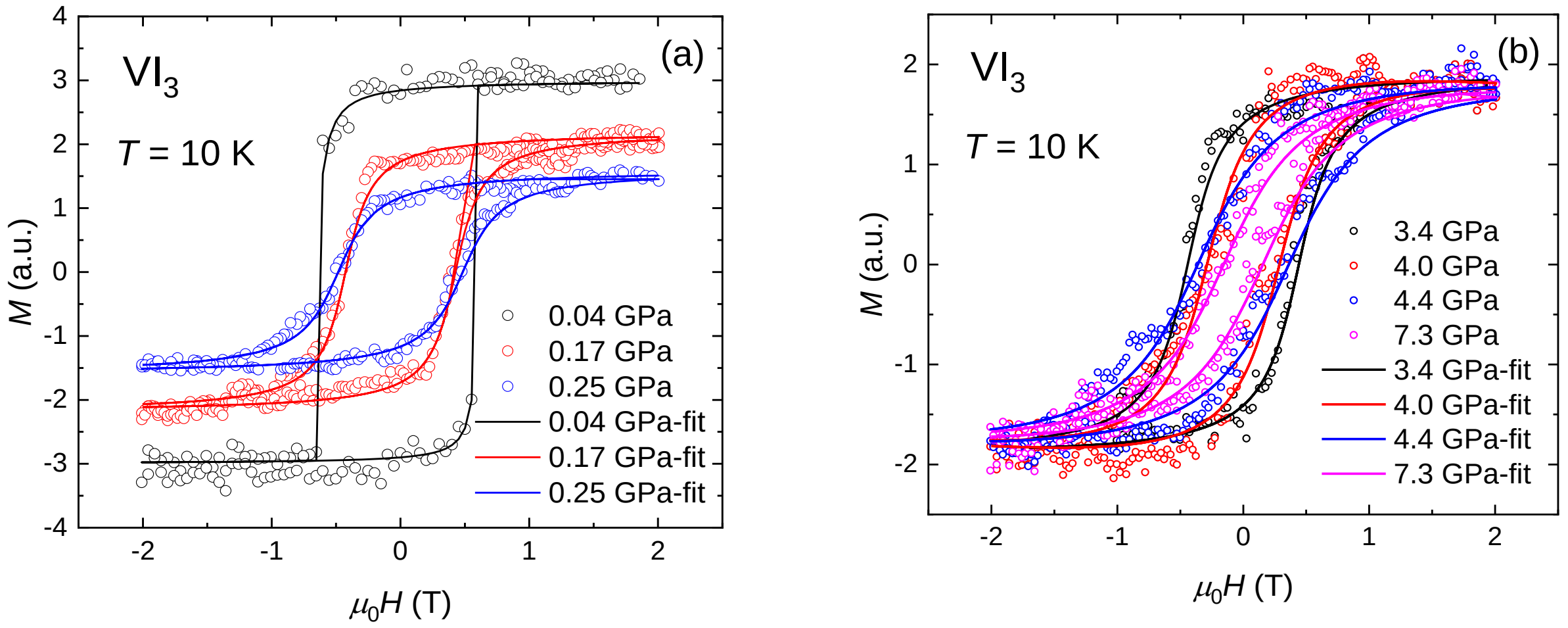

FIG. S7: Magnetization hysteresis loops of a $VI_3$ crystal measured at 10 K in the magnetic field applied along the *c**-axis at pressures: (a) 0.4, 1.7, and 2.5 GPa; (b) 3.4, 4.0, 4.4, and 7.3 GPa.

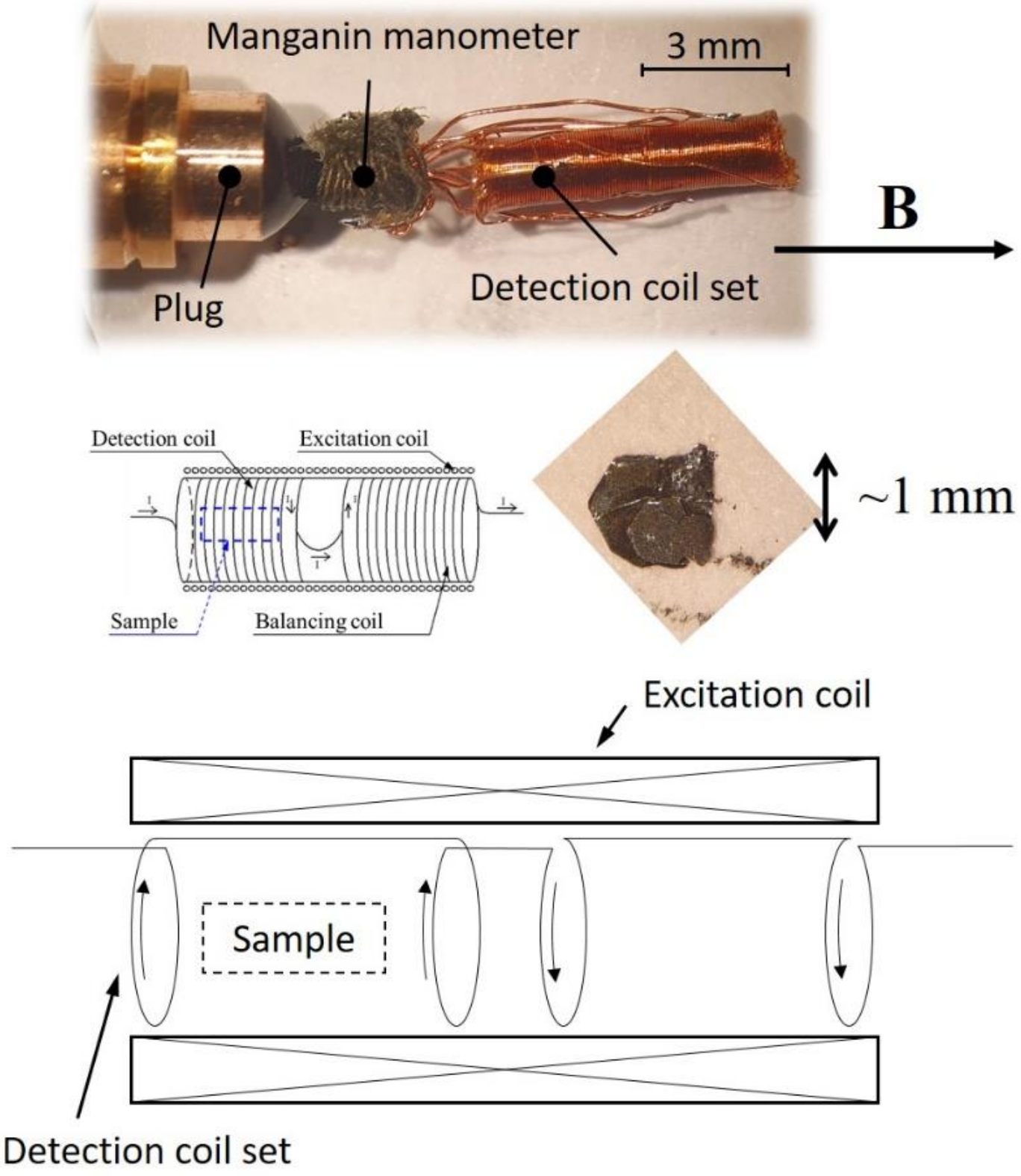

FIG. S8: The pressure experimental setup for measuring the AC susceptibility. Details of the pressure cell are described in the Experimental Details section.